\documentclass{article}
\usepackage{spconf,amsmath,graphicx}
\graphicspath{ {images/} }
\usepackage{multirow}
\usepackage{subcaption,stfloats}
\usepackage{enumitem}
\usepackage{xcolor}

\let\OLDthebibliography\thebibliography
\renewcommand\thebibliography[1]{
  \OLDthebibliography{#1}
  \setlength{\parskip}{0pt}
  \setlength{\itemsep}{3pt plus 0.3ex}
}

\title{A Principled Hierarchical Deep Learning Approach to Joint Image Compression and Classification}
\name{Siyu Qi, Achintha Wijesinghe, Lahiru D. Chamain, Zhi Ding\thanks{This material is based on works supported by the National Science Foundation under Grants 2002937 and 1824553.}}
\address{Department of Electrical and Computer Engineering \\
University of California Davis\\
Davis, CA, USA \\
\{syqi, achwijesinghe, hdchamain, zding\}@ucdavis.edu}

\begin{document}
%
\maketitle
\vspace*{-0.4in}
\begin{abstract}
Among applications of deep learning (DL) involving low cost sensors, remote image classification involves a physical channel that separates edge sensors and cloud classifiers. Traditional DL models must be divided between an encoder for the sensor and the decoder+classifier at the edge server. An important challenge is to effectively train such distributed models when the connecting channels have limited rate/capacity. Our goal is to 
optimize DL models such that the encoder latent requires low channel bandwidth while still delivers feature information for high classification accuracy. This work proposes a three-step joint learning strategy to guide encoders to extract features that are compact, discriminative, and amenable to common augmentations/transformations. We optimize latent dimension through an initial screening phase before end-to-end (E2E) training. To obtain an adjustable bit rate via a single pre-deployed encoder, we apply entropy-based quantization and/or manual truncation on the latent representations. Tests show that our proposed method achieves accuracy improvement of up to 1.5\% on CIFAR-10 and 3\% on CIFAR-100 over conventional E2E cross-entropy training.
\end{abstract}

\vspace*{-3mm}
\section{Introduction}
\vspace*{-1mm}
Recent advances have solidified
deep learning (DL) as a major tool 
in multimedia applications, including
compression and classification.
The rapid growth and broad deployment of low-cost sensing devices in Internet of Things (IoT) systems have further fueled the rise of distributed learning 
that relies on networked cooperation
of source and server nodes.
Within such learning framework, power- and memory-efficient encoding nodes are implemented to compress and transmit data to cloud/server nodes to implement subsequent DL tasks. 
In this work, we focus on AEs consisting of an encoder at the source node before the channel and classifier~\cite{chamain2022end,zhou2019CDSAE,chen2018cross} at the  channel output end, which turn out to be one of the foremost feasible solutions for bandlimited image classification. 
However, these AEs, trained by a rate loss accompanied by a weighted classification loss in an end-to-end (E2E) manner, may lead to a local minimum that neither brings down the coding rate nor optimizes the accuracy without careful manual tuning~\cite{qi2022hierarchical} and face three major challenges to be deployed in a networked environment.

Such AEs rely heavily on the sizable
training data.
In real life, training images may undergo various distortive transformations~\cite{rivenson2018deep,wang2022remote,guo2022jitter}, such as 
translation, mirroring, rotation, color jitter, etc. Therefore, encoders shall preserve distortion-invariant key information for classification . We handle this challenge by incorporating supervised learning and self-supervised learning (SSL) algorithm under the guidance of the principle of Maximal Coding Rate Reduction (MCR${}^2$)~\cite{yu2020mcr2}, which has been proven able to guide DL models to learn diverse and discriminative features from inputs.

In networked learning, bandwidth is an essential constraint to be considered by the encoders. As the transmission band limit can change in real-time, the encoders should provide the flexibility of various coding rates. We address this objective through two approaches: entropy-based quantization and latent truncation. Entropy-based quantization can be directly applied to any deployed encoders without further optimization. 
The other method, latent truncation, requires fine-tuning and updating the deployed encoders and decoders but provides greater bandwidth reductions while preserving the classification accuracy. These two methods can be used either individually or together on any pre-deployed AE.

The dimension of an AE's bottleneck layer is an essential parameter that impacts the compression ratio and classification accuracy. However, the intrinsic dimensionality of data varies between image datasets~\cite{ghojogh2019feature,bahadur2019dimension}, which requires expertise to determine when implementing an AE.
We tackle this issue via an initial CE-training screening process. 

In brief, the contributions of this paper are:
\vspace{-1mm}
\begin{enumerate}
    \item We implement a joint learning approach guided by the principle of MCR${}^2$. This approach can regularize the encoder to extract linear discriminant representations (LDRs) that are in-class compact, between-class discriminative and consistent to various augmentations/transformations.
    \vspace{-1mm}
    \item We propose two methods to obtain continuous latent bit rate via a single AE model: entropy-based quantization and manual latent truncation.
    \vspace{-1mm}
    \item We suggest adding an initial screening phase during the AE designing stage to select the optimal latent dimension.
\end{enumerate}
\vspace{-6mm}
\section{Related Work}
\vspace{-1mm}
\subsection{Linear Discriminative Representations (LDRs)}
\vspace{-1mm}
In networked learning scenarios, one way to effectively transmit high-dimensional real-world data is to map it to low-dimensional subspaces, which shall be linear in the ideal case. For image classification specifically, the extracted latents should be inter-class diverse, in-class discriminative, representative of the original data and robust against noise including real-life data distortion or channel errors.

Mathematically, one can denote an image dataset consisting of $N$ samples belonging to $[K]$ classes as $\boldsymbol{X} = \{\boldsymbol{x}_1, \boldsymbol{x}_2, \dots, \boldsymbol{x}_N\} \in \mathcal{R}^{D_{\rm in}\times N}$, with $D_{\rm in}$ being the input dimension. Each image has a class label $c_i \in [K], i= 1, \dots, N$. Conventionally, a deep classifier learns to map the input $\boldsymbol x$ to its label $c$ by CE loss. Based on the principle of MCR${}^2$, works in \cite{yu2020mcr2,ma2007segmentation} that this objective can be achieved by minimizing the novel LDR loss function:
\begin{equation}
\label{eq:mcr2_loss}
    \mathcal{L}_{\rm LDR}= -\Delta R(\boldsymbol Z) \dot{=} -R(\boldsymbol{Z})+R_c(\boldsymbol{Z}, \mathbf{\Pi}),
\end{equation}
with $R(\boldsymbol{Z})=\frac{1}{2}\textrm{$\log \det$} \left(\boldsymbol{I}+\frac{D_{\rm in}}{\epsilon^2 N} \boldsymbol{Z}\boldsymbol{Z}^T\right)$ being the average bit rate of the latent $\boldsymbol{z}_i\in \mathcal{R}^{D_{z}}$ up to a precision $\epsilon$. 
 If a grouping partition $\mathbf{\Pi} = \{\mathbf{\Pi}_j\}_{1}^{K}$ of $K$ classes is known, where $\mathbf{\Pi}_j$ is a $N\times N$
 diagonal matrix with values ``1'' indicating samples belonging to the $j$-th class and ``0'' otherwise, the class-wise average bit rate of $\boldsymbol{z}_i$ can be computed by $R_c(\boldsymbol{Z}, \mathbf{\Pi})= \sum_{j=1}^K \frac{\textrm{tr}(\mathbf{\Pi}_j)}{2N} \textrm{$\log \det$}\left(\boldsymbol{I}+\frac{D_{\rm in}}{\epsilon^2 \textrm{tr}(\mathbf{\Pi}_j)}\boldsymbol{Z} \mathbf{\Pi}_j \boldsymbol{Z}^T\right)$.

From an information-theoretic perspective, minimizing Eq.~(\ref{eq:mcr2_loss}) corresponds
to maximizing the difference between the global and group-wise average coding rates of the dataset. 
As elaborated in~\cite{yu2020mcr2}, this MCR${}^2$
principle generates robust, diverse and discriminant latent features from the data for classification. We consider this principle to be
consistent with the goal of
optimizing latent dimension
for transmission over rate-constrained
channel for distributed classification. 

\vspace{-3mm}
\subsection{Auto-Encoders and Hierarchical Learning}
\vspace{-1mm}
AEs have been very successful in joint feature extraction and classification~\cite{yao2020deep,shao2020bottlenet,jankowski2020joint}.
However, existing DL models
are typically
trained in an E2E manner through na\"ive superposition of
multiple objective functions
without guidance
or constraints on latent representations. The Compact and Discriminative Stacked AE (CDSAE)~\cite{zhou2019CDSAE} imposes a diverse regularization and a local Fisher's discriminant regularization~\cite{fisher1936lda} on each layer for a compact and discriminative mapping. The CDSAE is accompanied by two-phase hierarchical training, in which the first phase aims at low-dimensional feature extraction to emphasize in-class similarity and between-class diversity, whereas the second phase targets E2E joint compression and classification. Similarly, the work of~\cite{qi2022hierarchical} suggests a Dual-Phase Hierarchical Learning (DuPHiL) strategy. The first phase of DuPHiL fixes the decoder to optimize the encoder by using the MCR${}^2$ loss via a side path. This phase trains the encoder to map input images to
a compact latent space for efficient compression while preserving the necessary information for subsequent classification. The second phase freezes the encoder after phase one and trains the decoder by using the CE loss to learn accurate classification based on the LDRs generated by the encoder. Both works prove that by assigning different sub-tasks to different modules, overall rate-accuracy trade-off is improved.

\vspace{-3mm}
\section{Proposed Methods}
\vspace{-1mm}
\subsection{Hierarchical Training Based on Joint Supervised and Self-Supervised Learning (SSL)}
\begin{figure}[htb]
	\begin{center}
		\includegraphics[width=\linewidth]{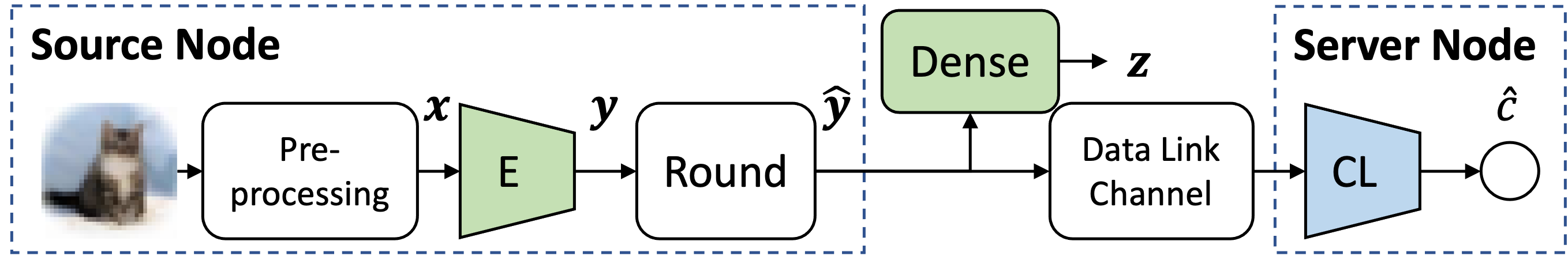}
	\end{center}
	\vspace{-3mm}
	\caption{Architecture of adopted AE. ``E'' denotes ``encoder'' and ``CL'' denotes ``classifier''.}
	\vspace{-4mm}
	\label{fig:architecture}
\end{figure}

\vspace{-1mm}
We adopt the AE architecture proposed in \cite{qi2022hierarchical} as shown in Fig.~\ref{fig:architecture}. The model consists of a pair of ResNet-based encoder and decoder as well as a lightweight side branch (for training only) from which the LDR loss can guide the encoder.

When sensor node captures real-life images, distortions are common~\cite{rivenson2018deep,wang2022remote,guo2022jitter}. 
The encoder should learn to extract the underlying general information to account for such
possible distortions with minimum loss. Self-supervised representation learning \cite{oord2018representation,he2020momentum,wu2018unsupervised,zhai2019s4l,kolesnikov2019revisiting,misra2020self} can train the encoders to extract valuable features for a downstream task and have demonstrated encouraging results. More specifically,
the work of \cite{yu2020mcr2} suggests that SSL with MCR${}^2$ principle is a promising approach to promoting LDRs' consistency 
under certain transformations/augmentations. Following~\cite{yu2020mcr2}, we
also augment each image $\mathbf{x}_i$ in a mini-batch
with $n$ transformations randomly drawn from a collection  $\mathcal{T}$ of augmentations with a know distribution $\mathcal{P}_\mathcal{T}$. These
augmented images belong to the same class. We define the objective function of LDR-guided SSL as
\begin{equation}
\label{eq:selfsup_loss}
    \mathcal{L}_{\rm LDR-SSL}= -\Delta R(\boldsymbol Z) \dot{=} -R(\boldsymbol{Z})+R_c(\boldsymbol{Z}, \mathbf{\Pi^\mathcal{T}}),
\end{equation}
where $\mathbf{\Pi^{\mathcal{T}}}$ is the artificial self-labeled partition matrix. 

However, model training by SSL algorithms alone has two main drawbacks. Firstly, the larger dataset
generated by multiple augmentations may lead to
overfitting on label-dependent but not feature-dependent
information~\cite{zhai2019s4l} and slower convergence. Secondly, SSL relies only on artificially-constructed labels instead of categorical ground truth labels, which are available in our framework, and usually performs poorer than supervised learning~\cite{zhai2019s4l,su2020does}. Therefore, we propose to incorporate supervised learning with SSL, as a feature regularizer, to jointly balance training speed, rate-accuracy performance and LDRs' robustness to distortion. To summarize, our three-step hierarchical training includes:
\vspace{-1mm}
\begin{enumerate}[leftmargin=*,label={Step \arabic*:}]
    \item Apply DuPHiL with Encoder loss $\mathcal{L}_{\rm LDR}$ and Decoder loss $\mathcal{L}_{CE}$, given the ground truth partition $\mathbf{\Pi}$.
    \vspace{-1mm}
    \item Apply DuPHiL with Encoder loss $\mathcal{L}_{\rm LDR-SSL}$ and Decoder loss $\mathcal{L}_{CE}$, given the artificial partition $\mathbf{\Pi^{\mathcal{T}}}$.
    \vspace{-1mm}
    \item Apply E2E training with loss $\mathcal{L}_{CE}$, while at a smaller learning rate than Steps 1 and 2.
\end{enumerate}

\vspace{-3mm}
\subsection{Selecting Latent Dimensions}
\vspace{-1mm}
Our tests show that an initial quick CE-training phase can guide the selection of latent dimension from a finite set of values for our AE model. This initial screening consists of training separate AEs of various latent dimensions to minimize CE loss before selecting one with the highest accuracy.

\vspace{-3mm}
\subsection{Entropy-Based Latent Quantization}
\vspace{-1mm}
\label{sect:method_entropy_quantization}
Recent work~\cite{cheng2020learned} attempt to obtain different encoding data rates by training
various AEs to manually adjusting a rate-accuracy trade-off parameter $\lambda$ in the joint-compression-and-classification loss $\mathcal{L} = \mathcal{L}_{R}+\lambda\mathcal{L}_{CE}$. Such a na\"ive loss
function, however, may lead to a
convergence to a compromising local minimum that neither
minimizes classification error nor reduces
the rate. Furthermore,
tuning $\lambda$ can be time-consuming and costly. 
In our work, as the compact latent $\mathbf{\hat{y}}$ naturally contains discriminative information, we suggest to train only one AE and reduce the data rate of latent $\mathbf{\hat{y}}$ by adjusting the quantization step sizes.

Since different entries in latent representations may carry
different levels of information (entropy), we assign a quantization step size $\boldsymbol{q}_i$ to each entry $ \mathbf{y}_{i} $, where $ i = 1,\ 2,\ \dots,\ d_\mathbf{y} $. And $\boldsymbol{q}_i$ is linearly proportional to average entropy $ H(\mathbf{y}_{i}) $ on the training set for Gaussian $ \mathbf{y}_{i} $. To be specific, $\boldsymbol{q}_i = s \times H(\mathbf{y}_{i}) $, where $s$ is a scalar adjusting the overall quantization level. We obtain the quantized latent $ \mathbf{\hat{y}}_i $ for subsequent encoding and transmission by $\mathbf{\hat{y}}_{i} = \frac{\mathbf{y}_{i}}{\boldsymbol{q}_i} = \frac{\mathbf{y}_{i}}{s \times H(\mathbf{y}_{i})}$.

Clearly, a smaller $ \boldsymbol{q}_{i} $ leads to smaller range of quantized latent values $ \mathbf{\hat{y}}_{i} $, thereby  generating fewer encoded bits
(i.e. lower rate). Thus, those entries with higher entropy (i.e., higher variations) are quantized with finer resolution. 

\vspace{-3mm}
\subsection{Manual Truncation on LDRs}
\vspace{-1mm}
Channel bandwidth can vary over time and the encoder needs to be updated accordingly in response. As the proposed entropy-based quantization is applied directly to latents without optimizing the encoder, it can yield sub-optimal rate-accuracy performance with large quantization step sizes $\boldsymbol{q}_i$. As an alternate way to reduce the coding rate of latent representations while achieving a good rate-accuracy trade-off, we propose to truncate the bottleneck layer in the encoder, which is equivalent to manually truncating certain entries in $\boldsymbol{\hat{y}}$, followed by decoder fine-tuning. 
Since both the overall diversity and in-class compactness of latents are fortified by LDR-guided fine-tuning, they are expected to be more robust against such truncation. 

\vspace{-3mm}
\section{Experimental Results}
\vspace{-1mm}
We train the AEs on CIFAR-10 and CIFAR-100 datasets, using a ResNet-18 and a ResNet-34 backbone architecture, respectively. To determine latent dimensions, we form AEs with a bottleneck layer dimension $d_\mathbf{y} \in \{32, 64, 128, 256, 512\}$, with side branch detached, and pre-train each model using CE loss for 200 epochs. After this training step, we compare the rate-accuracy performances to choose the best latent dimension and the selected model is labeled as the ``CE-trained'' (CE-T) baseline model. Using the pre-trained baseline model, we further apply DuPHiL~\cite{qi2022hierarchical} or our three-step training algorithm to generate the ``DuPHiL'' or ``LDR-fine-tuned'' (LDR-FT) models, respectively.

\vspace{-3mm}
\subsection{Rate-Accuracy Performance}
\begin{figure}[ht]
	\begin{center}
		\includegraphics[trim=7 5 0 7,clip,width=0.46\textwidth]{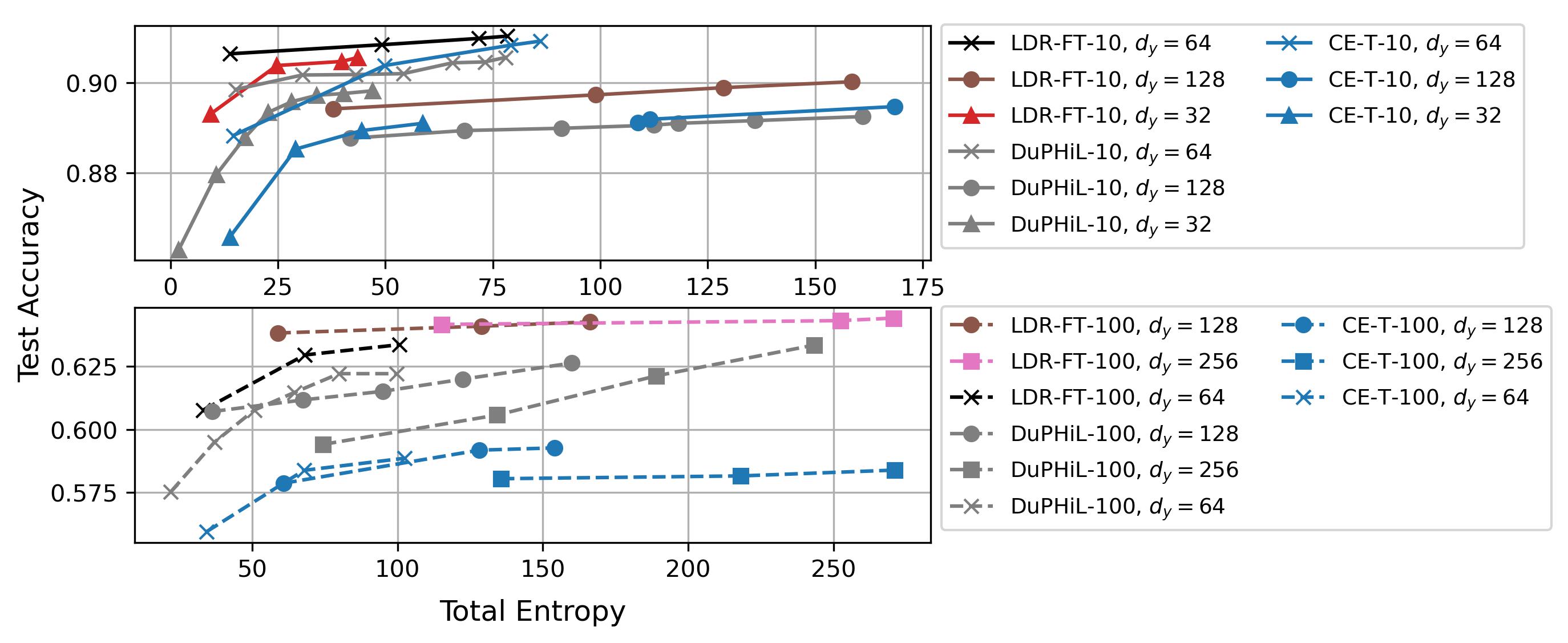}
    \vspace{-2mm}
	\caption{Accuracy of proposed LDR-FT, DuPHiL and CE-T models for CIFAR-10 or CIFAR-100.}
	\label{fig:acc_vs_entropy}
    \end{center}
\vspace{-5mm}
\end{figure}

\vspace{-1mm}
\label{sect:results_rate_acc}
We present the rate-accuracy performance of models trained by the proposed three-step training strategy (LDR-FT) together with the baseline  CE-T models, 
respectively, on CIFAR-10 with $d_\mathbf{y} \in \{32, 64, 128\}$ and CIFAR-100 with $d_\mathbf{y} \in \{64, 128, 256\}$ in Figs.~\ref{fig:acc_vs_entropy} and~\ref{fig:acc_vs_entropy_truncated}. 
Assuming each individual entry in the quantized latent representations $\mathbf{\hat{y}}$ follows Normal distribution,
the total entropy of $\mathbf{\hat{y}}$ is evaluated on the test set. To vary the entropy of each model, as discussed in Section~\ref{sect:method_entropy_quantization}, we assign a quantization step size $\mathbf{q}_i$ to each entry in $\mathbf{y}_i$ based on their entropy value and linearly scale the step size by a constant factor $s$. 

It is obvious that the classification performance correlates positively  
with total entropy. The LDR-FT models achieves a higher classification accuracy at the same total entropy compared with their corresponding CE and DuPHiL baselines, yielding a better entropy-accuracy trade-off. For example, with $d_\mathbf{y} = 64$ as the optimized latent dimension according to our experiments, 
the test accuracy achieved by 
baseline CE-T is $90.92\%$ whereas the LDR-FT model accuracy is
is $0.11\%$ higher at $91.03\%$. The LDR-FT model is also more robust to entropy-based quantization as its test accuracy drops by only $0.39\%$ when the total entropy is reduced from $71.68$ to $13.93$, whereas the accuracy of the baseline CE-T model decreases by $2.09\%$ when the total entropy is reduced from $86.02$ to $14.64$. Similarly, on CIFAR-100, at the same entropy, LDR-FT models usually outperform CE-T or DuPHiL in test accuracy. Further, the results also reveal unexpectedly that a higher latent dimension $d_\mathbf{y}$ does not always provide better classification accuracy, despite the ability to pack more information. We investigate and discuss this phenomenon in Section~\ref{sect:latent_dim}.

\vspace{-3mm}
\subsection{Robustness against Distortions and Truncation}
\begin{figure}[ht]
	\begin{center}
		\includegraphics[trim=7 5 0 7,clip,width=0.4\textwidth]{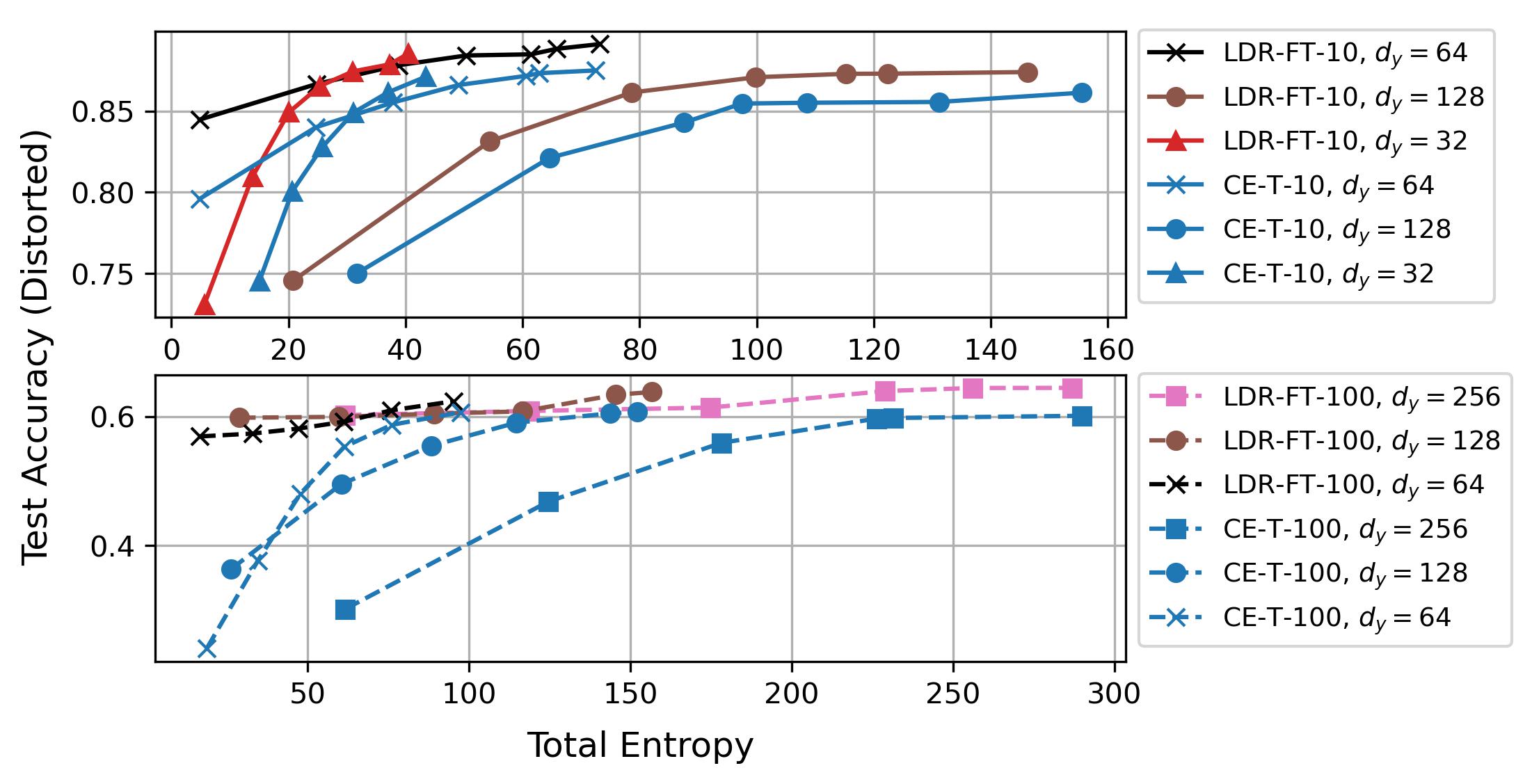}
  	\vspace{-2mm}
	\caption{Accuracy of LDR-FT versus CE-T models for distorted CIFAR-10 or CIFAR-100.}
	\label{fig:aug_test_acc_vs_entropy}
		\end{center}
  	\vspace{-5mm}
\end{figure}

\vspace{-1mm}
To illustrate the robustness against random distortions, we apply rotation, flipping and color jitter to the test sets and present the evaluation accuracy in Fig.~\ref{fig:aug_test_acc_vs_entropy}. Clearly, LDR-FT models achieve better accuracy, indicating that they can preserve the invariant key information for classification.

To illustrate the robustness against truncation when channel bandwidth is lowered, we manually truncate the last half latent entries in each of the same CE-T base models, before further fine-tuning them with either proposed training strategy (LDR-FT)  or E2E CE training (CE-T). 

\begin{figure}[ht]
	\begin{center}
		\includegraphics[trim=7 5 0 7,clip,width=0.4\textwidth]{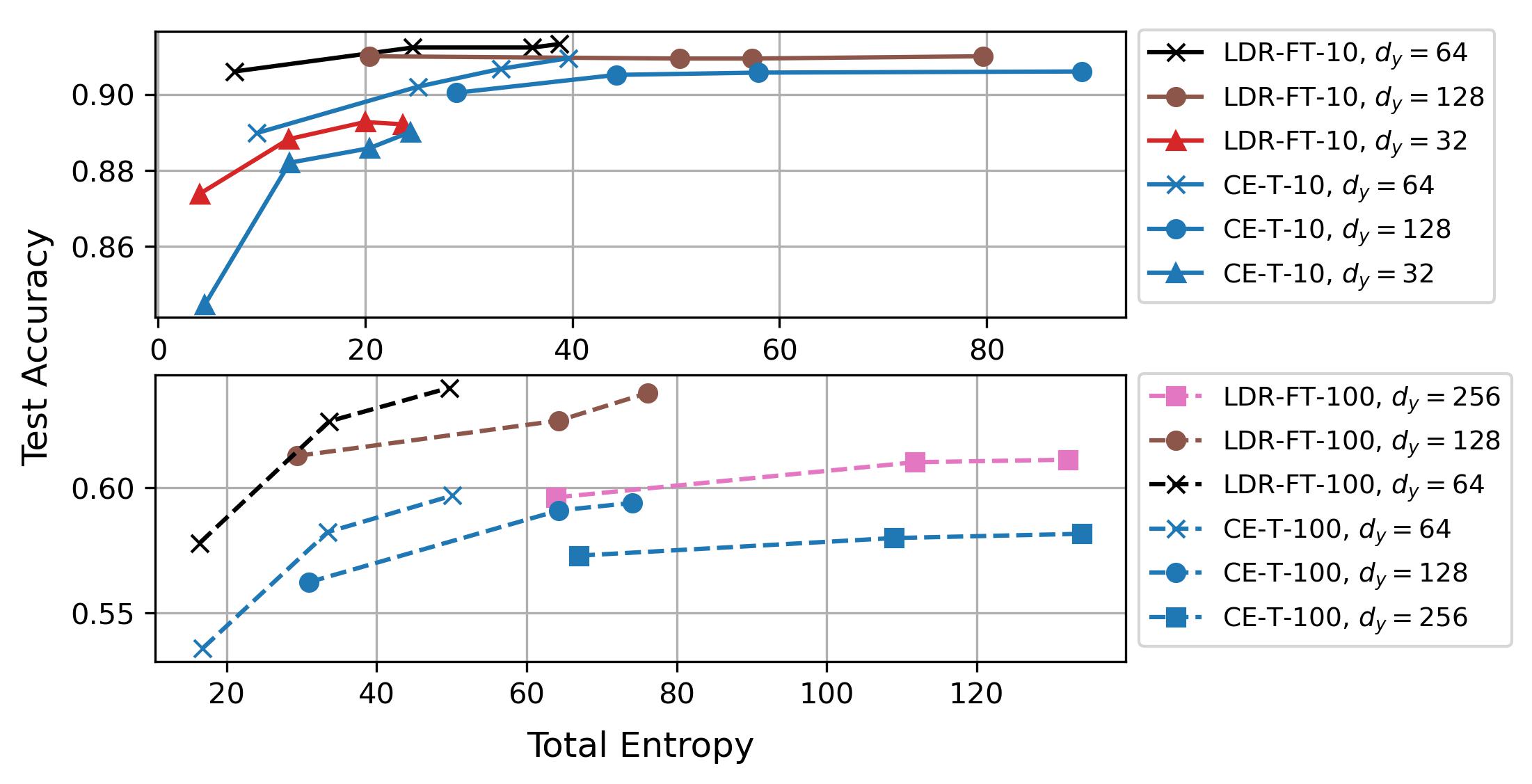}
  	\vspace{-2mm}
	\caption{Accuracy of LDR-FT versus CE-T models (with latent truncation) for CIFAR-10 or CIFAR-100.}
	\label{fig:acc_vs_entropy_truncated}
		\end{center}
  	\vspace{-5mm}
\end{figure} 

As shown in Figure~\ref{fig:acc_vs_entropy_truncated}, the truncated LDR-FT models outperform their
CE-T counterparts in terms of the entropy-accuracy trade-off, indicating that the LDR-FT training strategy fortifies models to be more robust against such active truncation in response to the lower channel bandwidth. Moreover, comparing the performance of $d_\mathbf{y} = 128$ models on CIFAR-10 before and after manual truncation in Figs.~\ref{fig:acc_vs_entropy} and~\ref{fig:acc_vs_entropy_truncated}, we can observe an accuracy improvement of approximately $1\%$, which again confirms our claim in Section~\ref{sect:results_rate_acc} that higher latent dimension is not necessarily always the optimal choice.

\vspace{-3mm}
\subsection{Impact of Latent Dimensions on Model Performance}
\vspace{-1mm}
\label{sect:latent_dim}
Conceptually, with a higher latent dimension, we expect better classification accuracy as the latent representations are more informative. However, experiments show that this does not always hold.  

\begin{figure}[ht]
	\begin{center}
		\includegraphics[trim=7 5 0 7,clip,width=0.4\textwidth]{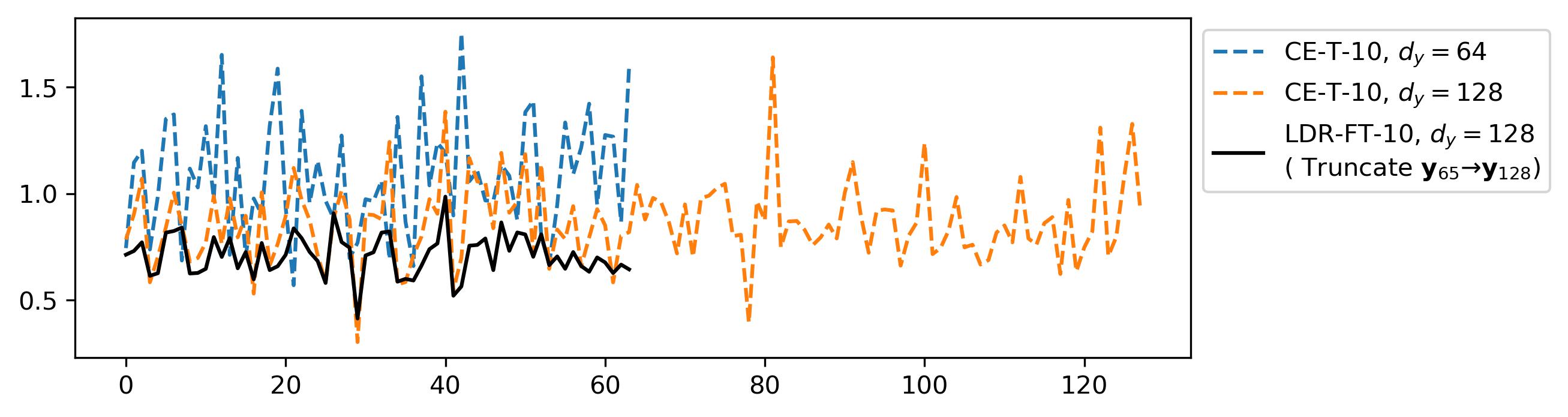}
  	\vspace{-2mm}
	\caption{Standard deviation of  entries in latent representation $\hat{\mathbf{Y}}$ computed on CIFAR-10 test set, with $d_\mathbf{y} = 64$ and $128$.}
	\label{fig:y_hat_std}
	\end{center}
 \vspace{-5mm}
\end{figure} 

To investigate the possible causes, without the loss of generosity, we examine the variations in the learned latent representations for dimension sizes $d_\mathbf{y} = 64$ and $128$, using the CIFAR-10 test set. As Fig.~\ref{fig:y_hat_std}
shows, comparing the two CE-T cases, the model with $d_\mathbf{y} = 64$ learns to pack more information in the latent by increasing variations, making its latent more robust against quantization noise in comparison with $d_\mathbf{y} = 128$. Meanwhile, after applying manual truncation and fine-tuning ``CE-T-10, $d_\mathbf{y} = 128$'', the ``LDR-FT-10'' model
learns to suppress latent variations, leading to an entropy reduction while preserving the most critical between-class discriminative information, thereby avoiding classification accuracy loss as can be seen in Fig.~\ref{fig:acc_vs_entropy}.

\vspace{-3mm}
\section{Conclusions}
\vspace{-1mm}
This work studies the
training of AE for distributed 
compression or classification
in distributed learning environment. We 
propose a
combination of supervised and self-supervised training algorithms, guided by the information theoretic LDR criterion. 
To adjust transmission rate 
in response to channel bandwidth, we propose an entropy-based quantization, which operates directly on any pre-trained encoders, and we propose
a simple latent truncation
in conjunction with encoder fine-tuning. Our proposed
learning strategies can apply directly to various existing AE architectures for a compact, discriminant and distortion-robust latent mapping. Further, we investigate the impact of latent dimensions on the performance of AEs and suggest to optimize latent dimensions by using an initial screening process.

\bibliographystyle{IEEEtran}
\bibliography{reference}

\end{document}